\documentclass[aps]{article}

\begin{document}
\title{Neutrino Propagation and Oscillations in a Strong Magnetic
Field\footnote{%
Talk given at the Conference: MRST-2004: "From Quaks to
Cosmology," Montreal, Canada, May 12-14, 2004 }}
\author{Efrain J. Ferrer and Vivian de la Incera}
\date{Physics Department, State University of New York at Fredonia,
Fredonia, NY 14063, USA}
\maketitle
\begin{abstract}
We review the results on neutrino propagation in neutral and
charged media under strong magnetic fields $M_{W}^{2}\gg B\gg
m_{e}^{2}$. It is shown that the neutrino energy density gets a
magnetic contribution in both charged and neutral media, which is
linear in the magnetic field, of first order in $G_{F}$, and
independent of the charge density. This new term enters as a
correction to the neutrino kinetic energy and produces an
anisotropic contribution to the neutrino index of refraction. As a
consequence, in a neutral medium a highly anisotropic resonant
level-crossing condition takes place for the oscillation between
electron-neutrinos and the other neutrino species. Possible
cosmological applications are presented.
\end{abstract}

\section{Introduction}

In our universe the magnetic field is omnipresent. Large magnetic fields $%
B=10^{12}-10^{14}$ $G$ have been associated with the surface of
supernovas \cite{Ginzburg} and neutron stars \cite{Fushiki}, and
fields perhaps as large as $10^{16}$ $G$ with magnetars
\cite{Duncan}. Even larger fields could exist in the star's
interior. Based on the scalar virial theorem \cite{Lai}, it has
been suggested that the interior field in neutron stars could be
as high as $10^{18}$ $G$. On the other hand, the early universe
was presumably permeated by large primordial magnetic fields that
eventually became the seeds \cite{Grasso,enq} of the galactic
fields that are observed today at scales of $100$ Kpc and larger
\cite{Kronberg}.

In cosmology, the effects of primordial magnetic fields on
neutrino propagation are of special interest during the neutrino
decoupling era, since in that epoch a large number of these
particles escaped the electroweak plasma with the possibility to
develop significant oscillations. As for astrophysics, the
interconnection between a star's magnetic field and its particle
current flow, is of great relevance to understand the star's
evolution.

Thus, a thorough study of the effects of strong magnetic fields on
neutrino propagation is crucial for both astrophysics and
cosmology. In this sense it should be noticed that unlike the
stellar material, whose density can be of order one, the early
universe was almost
charge symmetric, with a particle-antiparticle asymmetry of only $%
10^{-9}-10^{-10}$. Therefore, when considering possible
applications to the early universe, we should restrict the
discussion to the effects on a magnetized neutral medium ($\mu
=0$, where $\mu $ is the chemical potential characterizing the
system's unbalance between particles and antiparticles); while in
the interior of neutron stars it is natural to consider a
magnetized charged medium ($\mu \neq 0$).

The propagation of neutrinos in magnetized media has been
previously investigated by several authors [8-11]. The weak-field
results of papers [8-11] led to think that the magnetic-field
effects could be significant in astrophysics, because of the
field- and $\mu $-dependent terms of order $1/M_{W}^{2}$, but were
irrelevant ($\sim 1/M_{W}^{4}$-order) in the early universe due to
its charge-symmetric character ($\mu =0$). However, as we have
recently found [12-15], and will discuss below, a strong
magnetic field ($M_{W}\gg \sqrt{B}\gg m_{e},T,\mu ,\left| \mathbf{p}\right| $%
) gives rise to a new contribution to the neutrino energy that is
linear in the field, independent of the chemical potential, and
that is of the same order ($1/M_{W}^{2}$) as the largest terms
found in the weak-field charged-medium case. This new result turns
magnetic-field effects relevant for cosmology.

\section{Neutrino Self-Energy in Magnetized Media. Covariant Structure}

The neutrino self-energy operator $\sum (p)$ is a Lorentz scalar
that can be formed in the spinorial space taking the contractions
between the characteristic vectors and tensors of the system with
all the independent elements of the Dirac ring. Explicit chirality
reduces it to
\begin{equation}
\sum =R\overline{\sum }L,\qquad \overline{\sum }=V_{\mu }\gamma
^{\mu } \label{A1}
\end{equation}
where $L,R=\frac{1}{2}(1\pm \gamma _{5})$ are the chiral
projectors, and $V_{\mu }$ is a Lorentz vector that can be spanned
as a superposition of four basic vectors formed with the
characteristic tensors of the problem. In a magnetized medium,
besides the neutrino four-momentum $p_{\mu }$, and the
four-velocity of the center of mass of the magnetized medium
$u_{\mu },$ we have to consider the magnetic-field strength tensor
$F_{\mu \nu }$. In a neutral medium it is convenient to use the
expansion
\begin{equation}
\overline{\sum }(p,B)=ap\llap/_{\Vert }+bp\llap/_{\perp }+cp^{\mu }\widehat{%
\widetilde{F}}_{\mu \nu }\gamma ^{\nu }+idp^{\mu }\widehat{F}_{\mu
\nu }\gamma ^{\nu }.  \label{A2}
\end{equation}
where  we introduced the notations $\widehat{F}_{\mu \nu
}=\frac{1}{\left| B\right| }F_{\mu \nu }$, $\widehat{\widetilde{F}}_{\mu \nu }=$ $\frac{1}{%
2\left| B\right| }\varepsilon _{\mu \nu \rho \lambda }F^{\rho
\lambda }$, $p\llap/_{\Vert }=p^{\mu }\widehat{\widetilde{F}}_{\mu
}\,^{\rho }\widehat{\widetilde{F}}_{\rho \nu }\gamma ^{\nu },$ and
$p\llap/_{\perp }=p^{\mu }\widehat{F}_{\mu }\,^{\rho
}\widehat{F}_{\rho \nu }\gamma ^{\nu }.$

In a charged medium, it will be more convenient (the coefficients
in this
basis will be non-singular in the $p\rightarrow 0$ limit) to expand $\overline{%
\sum }$ in a different basis

\begin{equation}
\overline{\sum }(p,B)=a^{\prime }p\llap/+b^{\prime }u\llap/+c^{\prime }%
\widehat{B}\llap/+id^{\prime }p^{\mu }\widehat{F}_{\mu \nu }\gamma
^{\nu }. \label{A3.1}
\end{equation}

In the covariant representation the magnetic field is expressed as
$B_{\mu }=\frac{1}{2}\varepsilon _{\mu \nu \rho \lambda }u^{\nu
}F^{\rho \lambda }$. The coefficients $a$($a^{\prime }$),
$b$($b^{\prime }$), $c$($c^{\prime }$), and $d$($d^{\prime }$) are
Lorentz scalars that depend on the parameters of the theory and
the approximation used. Our goal is to find those coefficients in
the one-loop approximation, and hence, the dispersion relations
for different systems in the presence of strong magnetic-field
backgrounds.

\section{Neutrino Dispersion Relation in Strongly Magnetized Media}

The neutrino dispersion relation in a magnetic background can be
found as the nontrivial solution of
\begin{equation}
\det \left[ p\llap/+\sum (p,B)\right] =0.  \label{B1}
\end{equation}

In the one-loop approximation, the neutrino self-energy is given
by
\begin{equation}
\Sigma (x,y)=\frac{ig^{2}}{2}R\gamma _{\nu }S(x,y)\gamma ^{\mu
}G_{F}(x,y)_{\mu }\,^{\nu }L  \label{B2}
\end{equation}
where $S(x,y)$ and $G_{F}(x,y)_{\nu }\,^{\mu }$ are the Green's
functions of the electron and W-boson, in the presence of the
magnetic field, respectively. Since the two virtual particles are
electrically charged, the magnetic field interacts with both
producing a Landau quantization of the corresponding transverse
momenta [12-15].

In a medium the neutrino self-energy operator can be unfolded as
\begin{equation}
\overline{\Sigma }(p,B)=\overline{\Sigma }_{0}(p,B)+\overline{\Sigma }%
_{T}(p,B),  \label{B3}
\end{equation}
where $\overline{\Sigma }_{0}$ is the field-dependent vacuum
contribution and $\overline{\Sigma }_{T}$ is the statistical part
which is a function of the temperature and the magnetic field (in
a charged medium the statistical part
will also depend on the chemical potential and will be represented by $%
\overline{\Sigma }_{T,\mu }(p,B)$). Our interest is to find $\overline{%
\Sigma }(p,B)$ in the strong-field approximation defined by the
condition $M_{W}\gg \sqrt{B}\gg m_{e},T,\left| \mathbf{p}\right|
$. Since in this approximation the gap between the electron Landau
levels is larger than the rest of the parameters entering in the
electron energy, it is consistent to use the lowest Landau level
(LLL) approximation in the electron Green's function, while in the
W-boson Green's function, because $M_{W}\gg \sqrt{B},$ we must sum
in all W-boson Landau levels.

\subsection{Neutral Case}\label{subsec:prod}

Within the strong field approximation, the vacuum and statistical
parts of
the neutrino self-energy in a magnetized neutral medium were found in Refs.~%
[12], and [15] respectively. Up to leading order in powers of
$1/M_{W}^{2}$ it is given by

\begin{equation}
\overline{\Sigma }_{T}(p,B)=ap\llap/_{\Vert }+cp^{\mu }\widehat{\widetilde{F}%
}_{\mu \nu }\gamma ^{\nu },  \label{B4}
\end{equation}

\begin{equation}
a=-c=\frac{-g^{2}eB}{8M_{W}^{2}}[\frac{1}{(2\pi )^{2}}+\frac{T^{2}}{%
3M_{W}^{2}}]\exp (-p_{\perp }^{2}/eB).  \label{B4.4}
\end{equation}

Solving the dispersion equation (\ref{B1}) for the self-energy
(\ref{B4}) we obtain the neutrino energy spectrum in the strongly
magnetized neutral medium as

\begin{equation}
E_{\pm }=\left| \mathbf{p}\right| [1-a\sin ^{2}\alpha ] \label{B5}
\end{equation}
In (\ref{B5})$,\alpha $ is the angle between the direction of the
neutrino momentum and that of the applied magnetic field, and
$E_{\pm }\equiv \pm p_{0}$ are the neutrino ($+$)/anti-neutrino
($-$) energies, respectively.

From (\ref{B4.4}) it is clear that in the strong-field limit the
thermal correction to the neutrino energy is much smaller ($\sim
1/M_{W}^{2}$ smaller) than the vacuum correction. This result can
be significant for cosmological applications in case a strong
primordial magnetic field were present in the primeval plasma. The
neutrino energy spectrum (\ref{B5}) depends on the neutrino
direction of motion, with a maximum field effect for neutrinos
propagating perpendicularly to the magnetic field.

\subsection{Charged Case}\label{subsec:prod}

We consider now a medium with a finite density of electric charge.
As usual, a finite density is reflected through the introduction
of a chemical potential $\mu $, which plays the role of an
external parameter. Here the following comment is in order, in
electroweak matter at finite density there exists the possibility
to induce additional ``chemical potentials'' \cite{Shabad}. These
chemical potentials are nothing but dynamically generated
background
fields given by the average $\left\langle W_{0}^{3}\right\rangle $ and $%
\left\langle B_{0}\right\rangle $ of the zero components of the
gauge fields. They are known to appear, for instance, in the
presence of finite lepton/baryon density \cite{Shabad} (for recent
applications related to this effect see Refs.~[17]). In our case,
a possible consequence of the condensation of such average fields
could be the modification of the effective chemical potential of
the W-boson, which then might be different from the electron
chemical potential. Nevertheless, that modification, even if
present in the case under study, will have no relevant
consequences,
as the W-boson contribution will always be suppressed by the factor $%
e^{-M_{W}/T}$.

We find \cite{Vivian} that the leading contribution to the
neutrino self-energy in a strongly magnetized charged medium is
\begin{equation}
\overline{\Sigma }_{T,\mu }(p,B)=b^{\prime }u\llap / +c^{\prime }\widehat{B}%
\llap/  \label{B6}
\end{equation}
\begin{equation}
b^{\prime }=-c^{\prime }=\frac{g^{2}eB\mu }{2(4\pi )^{2}M_{W}^{2}}%
e^{-p_{\bot }^{2}/2eB}[p_{0}-p_{3}-4\mu ]  \label{B6.6}
\end{equation}

Using the result (\ref{B6}) to solve the dispersion relation
(\ref{B1}), we find that the neutrino energy spectrum in the
strongly magnetized charged medium is given in leading order in
the Fermi coupling constant by

\begin{equation}
E_{\pm }=\left| \mathbf{p}\right| [1-a(T=0)\sin ^{2}\alpha ]-\mathbf{%
\mathcal{M}.B}\pm \mathnormal{E_{0}}  \label{B7}
\end{equation}
where
\begin{equation}
\mathcal{M}\equiv \frac{-\mathnormal{E_{0}}}{\left| \mathbf{B}\right| }\frac{%
\mathbf{p}}{\left| \mathbf{p}\right| },\qquad \
E_{0}=\sqrt{2}G_{F}e^{-p_{\bot
}^{2}/2eB}[N_{0}^{-}-N_{0}^{+}].\label{B9}
\end{equation}
and we introduced the notation $N_{0}^{\mp }$ for the
electron/positron number densities in the LLL.

In the RHS of Eq. (\ref{B7}), the first term is the modified
neutrino kinetic energy, with the field-dependent radiative
correction, $a(T=0)$, obtained after taking $T=0$ in Eq.
(\ref{B4.4}); the second term can be interpreted as a
magnetic-moment/magnetic-field interaction energy, with
$\mathcal{M}$ playing the role of a neutrino effective magnetic
moment; and the third term is a rest energy induced by the
magnetized charged medium. In the case under study we have that
although the neutrino is a neutral massless particle, the charged
medium can endow it, through quantum corrections, with a magnetic
moment proportional to the induced neutrino rest energy.

In a charged medium CP-symmetry is violated \cite{Pal}. A common
property of electroweak media with CP violation is the separation
between neutrino and anti-neutrino energies. In the present case
this property is manifested in the double sign of $E_{0}$ in
(\ref{B7}).

\section{Neutrino Oscillations in a Strongly Magnetized Primeval Plasma}

The strength of the primordial magnetic field in the neutrino
decoupling era can be estimated from the constraints derived from
the successful nucleosynthesis prediction of primordial $^{4}He$
abundance \cite{He}$,$ as well as from neutrino mass and
oscillation limits \cite{Enqvist}. These constraints, together
with the energy equipartition principle, lead to the relations
\begin{equation}
m_{e}^{2}\leq eB\leq m_{\mu }^{2},\qquad B/T^{2}\sim 2, \label{C1}
\end{equation}
with $m_{\mu }$ being the muon mass. It is reasonable to assume
the existence of a primordial magnetic field in the above range in
the epoch between the QCD phase transition and the end of
nucleosynthesis \cite {Elizalde}.

Notice that a field satisfying (\ref{C1}) would be effectively
strong with respect to the electron-neutrino, but weak for the
remaining neutrino flavors. If such a field existed during the
decoupling era, it could significantly affect the $\nu
_{e}\leftrightarrow \nu _{\mu },\nu _{\tau }$
and $\nu _{e}\leftrightarrow \nu _{s}$ resonant oscillations \cite{Australia}%
, as the field would differently modify the energy of $\nu _{e}$
compared to those of $\nu _{\mu }$, $\nu _{\tau }$ and $\nu _{s}.$

Let us consider the evolution equation in the presence of a strong
magnetic field for a two-level system
\begin{equation}
\frac{d}{dt}\left(
\begin{array}{l}
\nu _{e} \\
\nu _{\mu }
\end{array}
\right) =H_{B}\left(
\begin{array}{l}
\nu _{e} \\
\nu _{\mu }
\end{array}
\right)  \label{14}
\end{equation}
where the Hamiltonian $H_{B}$ is given by

\begin{equation}
H_{B}=p+\frac{m_{1}^{2}+m_{2}^{2}}{4p}+\left(
\begin{array}{ll}
-\frac{\Delta m^{2}}{4p}\cos 2\theta +E_{\pm} & \frac{\Delta
m^{2}}{4p}\sin
2\theta  \\
\frac{\Delta m^{2}}{4p}\sin 2\theta  & \frac{\Delta m^{2}}{4p}\cos
2\theta
\end{array}
\right)   \label{15}
\end{equation}
Here, $\theta $ is the  vacuum mixing angle, $\Delta
m^{2}=m_{2}^{2}-m_{1}^{2}$ is the mass square difference of the
two mass eigenstates, and the magnetic energy density contribution
to the electron-neutrino $E_{\pm}$ is given in Eq. (\ref{B5}).

In Eq. (\ref{15}) the magnetic field contribution to the
muon-neutrino in the second diagonal term has been neglected,
taking into account that it will be of second order in the Fermi
coupling constant as corresponds to the weak-field approximation
in the neutral medium.

If we consider that the only flavor present at the initial time
was the electron-neutrino, using the evolution equation
(\ref{14})-(\ref{15}) we find that the appearance probability for
the muon neutrino is given by

\begin{equation}
P_{B}\left( \nu _{e}\rightarrow \nu _{\mu }\right) =\sin
^{2}\theta _{B}\sin ^{2}\frac{\pi x}{\lambda }  \label{20}
\end{equation}
where $\lambda $ is the oscillation length in the magnetized space

\begin{equation}
\lambda =\frac{\lambda _{0}}{[\sin ^{2}2\theta +(\cos 2\theta
-\frac{\lambda _{0}}{\lambda _{e}})]^{1/2}}  \label{21}
\end{equation}
and

\begin{equation}
\sin ^{2}2\theta _{B}=\frac{\sin ^{2}2\theta }{(\cos 2\theta
-\frac{\lambda _{0}}{\lambda _{e}})^{2}+\sin ^{2}2\theta }
\label{23}
\end{equation}
is the probability amplitude written in terms of the vacuum
($\lambda _{0}=\frac{4\pi p}{\Delta
m^{2}}$) and magnetic ($\lambda _{e}=\frac{2\pi }{%
E_{\pm}}$) oscillation lengths.

If the resonant condition

\begin{equation}
\frac{\lambda _{0}}{\lambda _{e}}=\cos 2\theta   \label{24}
\end{equation}
is satisfied, then the probability amplitude (\ref{23}) will get
is maximum value independently of the value of the mixing angle in
vacuum $\theta $  (the same effect will be obtained between
electron-neutrinos and tau or sterile neutrinos). The condition
(\ref{24}) is a resonant level-crossing condition. We underline
that the resonant effect in the magnetized neutral medium is
anisotropic, as it depends on the direction of propagation of the
electron-neutrino with
respect to the magnetic field (see that $\lambda _{e}$ depends on $\alpha $%
).

The resonant phenomenon here is similar to that in the well known
MSW effect \cite{msw} in a charged medium. Nevertheless, we stress
that in the magnetized neutral medium the oscillation process does
not differentiate between neutrinos and antineutrinos.

\section{Concluding Remarks}

A main outcome of our investigation was to show that in strongly
magnetized systems a term of different nature emerges in both
charged and neutral media. The new term, which is linear in the
magnetic field and of first order in $G_{F}$, enters as a
correction to the neutrino kinetic energy. This correction is
present even in a strongly magnetized vacuum \cite{Elizalde},
since it is related to the vacuum part ($T=0,$ $\mu =0$) of the
neutrino self-energy at $B\neq 0$. A characteristic of the
field-dependent corrections to the neutrino energy is that they
produce an anisotropic index of refraction. We call attention that
the anisotropy does not differentiate between neutrinos and
antineutrinos.

The charged-medium results reported in the current work can be of
interest for the astrophysics of neutrinos in stars with large
magnetic fields. On the other hand, our finding for the neutral
medium can have applications in cosmology, if the existence of
high primordial magnetic fields is finally confirmed. Contrary to
some authors' belief \cite{Weak-F}, that regardless of the field
intensity, the neutrino dispersion relation in the early universe
is well approximated by the dispersion relation in the zero field
medium, our results indicate that if strong, and even
weakly-strong \cite{Vivian} magnetic fields existed in the
neutrino decoupling era, they could have an impact in neutrino
flavor oscillations in the primeval plasma.

\section{Acknowledgment}

This work was supported by NSF-grant PHY-0070986


\medskip

\begin{thebibliography}{}
\bibitem{Ginzburg}  V. Ginzburg, High Energy Gamma Ray Astrophysics
(North-Holland, Amsterdam, 1991); D. Bhattacharya and G.
Srinivasan, in X-ray Binaries, ed. W.H.G. Lewin, J. van Paradijs,
and E.P.J. van den Heuvel, Cambridge Univ. Press, Cambridge, 1995,
pag. 495.

\bibitem{Fushiki}  I. Fushiki, E. H. Gudmundsson, and C.J. Pethick,
Astrophys. J.\textbf{342}, 958 (1989); T.A. Mihara, et. al.,
Nature (London)
\textbf{346}, 250 (1990); G. Chanmugam, Annu. Rev. Astron. Astrophys. \textbf{%
30}, 143 (1992).

\bibitem{Duncan}  R. C. Duncan and C. Thompson, Astrophys. J. \textbf{392}
(1992) L9; B. Paczynski, Acta Astron. \textbf{42}, 145 (1992); V.
V. Usov, Nature \textbf{357}, 472 (1992); C. Thompson and R. C.
Duncan Astrophys. J. \textbf{473}, 322 (1996); C. Kouveliotou et
al., Nature \textbf{393}, 235 (1998); Astrophys. J. \textbf{510},
L115 (1999).

\bibitem{Lai}  D. Lai and S.L. Shapiro, Astrophys. J. \textbf{383}, 745 (1991)
and references therein.

\bibitem{Grasso}  D. Grasso and H.R. Rubinstein, Phys.
Rep. \textbf{348}, 163 (2001).

\bibitem{enq}  K. Enqvist, Int. J. Mod. Phys. D\textbf{7}, 331 (1998).

\bibitem{Kronberg}  Y. Sofue, M. Fujimoto, and R. Wielebinski, Ann. Rev.
Astron. Astrophys. \textbf{24}, 459 (1986); P. P. Kronberg, Rep. Prog. Phys. %
\textbf{57}, 325 (1994); R. Beck et. al., Ann. Rev. Astron. Astrophys. \textbf{%
34}, 153 (1996); R. Beck et. al., Ann. Rev. Astron. Astrophys.
\textbf{34}, 153 (1996).

\bibitem{McKeon}  G. McKeon, Phys. Rev. D\textbf{24}, 2744 (1981).

\bibitem{Feldman}  A. Erdas, and G. Feldman, Nucl. Phys.
B\textbf{343}, 597 (1990).

\bibitem{Weak-F}  V. B. Semikoz and J. W. F. Valle, Nucl. Phys.
B\textbf{425}, 651 (1994); P. Elmfors, D. Grasso and G. Raffelt,
Nucl. Phys. B\textbf{479}, 3 (1996); A. Erdas, C. W. Kim and T. H.
Lee, Phys. Rev. D\textbf{58}, 085016 (1998); D. Grasso, Nucl.
Phys. B (Proc. Suppl.) \textbf{70}, 267 (1999); A. Erdas, and C.
Isola, Phys. Lett. B\textbf{494}, 262 (2000).

\bibitem{Nieves}  J. C. D'Olivo, J. F. Nieves and P. B. Pal, Phys. Rev. D%
\textbf{40}, 3679 (1989).

\bibitem{Elizalde}  E. Elizalde, E. J. Ferrer and V. de la Incera, Ann. of
Phys. \textbf{295}, 33 (2002).

\bibitem{Efrain}  E. J. Ferrer, in Proceedings of the International
Conference ''Quantization, Gauge Theory, And Strings,'' 5-10 Jun
2000, Moscow, Russia (Scientif World Publi., Singapore 2001, V.
II, pag.301, editors A. Semikhatov, M. Vasiliev and V. Zaikin); E.
Elizalde, E. J. Ferrer and V. de la Incera, Neutrino Dispersion in
Intense Magnetic Field. Proceedings of Particle Physics and
Cosmology Second Tropical Workshop (AIP Conference Proceedings
V540, pag. 372, editor J. F. Nieves), San Juan, Puerto Rico, Apr.
2000.

\bibitem{Australia}  E. J. Ferrer and V. de la Incera, Neutrinos under
Strong Magnetic Fields. Proceedings of Neutrinos, Flavor Physics
and Precision Cosmology, Fourth Tropical Workshop (AIP Conference
Proceedings V689, pag. 372, editor J. F. Nieves and R. R. Volkas),
Cairns, Queensland, Australia, June 2003.

\bibitem{Vivian}  E. Elizalde, E. J. Ferrer and V. de la Incera,
hep-ph/0404234 (To appear in Phys. Rev. D\textbf{70} ).

\bibitem{Shabad}  A. D. Linde, Phys. Lett. B\textbf{86}, 39 (1979); E. J.
Ferrer, V. Incera and A. E. Shabad, Phys. Lett. B\textbf{185}, 407
(1987); Nucl. Phys. B\textbf{309}, 120 (1988).

\bibitem{Gusynin} A. Gynther, Phys. Rev. D\textbf{68}, 016001 (2003); F. Sannino and
K. Tuominen, Phys. Rev. D\textbf{68}, 016007 (2003); V. P.
Gusynin, V. A. Miransky and I. Shovkovy, Phys. Lett. B\textbf{58},
82 (2004); Mod. Phys. Lett. A\textbf{19}, 1341 (2004).

\bibitem{Pal}  P. B. Pal and T. N. Pham, Phys. Rev. D\textbf{40}, 259 (1989);
J. F. Nieves, Phys. Rev. D\textbf{40}, 866 (1989).

\bibitem{He}  K. A. Olive, D. N. Schramm, G. Steigman, and T. Walker, Phys.
Lett. B\textbf{236}, 454 (1990); L. M. Krauss and P. Romanelli,
Astrophys. J. \textbf{358}, 47 (1990).

\bibitem{Enqvist}  K. Enqvist, V. Semikoz, A. Shukurov, and D. Sokoloff,
Phys. Rev. D\textbf{48}, 4557 (1993).

\bibitem{msw}  L. Wolfenstein, Phys. Rev. D\textbf{17}, 2369 (1978); S. P.
Mkheyev, A. Yu. Smirnov, Sov. J. Nucl. Phys. \textbf{42}, 913
(1985).
\end{thebibliography}
\end{document}